\documentclass{article}
\usepackage{cite}
\usepackage{graphicx}
\usepackage{dcolumn}

\begin{document}

\date{}
\title{A pedagogical description of diabatic and adiabatic molecular processes}
\author{Francisco M. Fern\'{a}ndez \thanks{E-mail: fernande@quimica.unlp.edu.ar}\\
INIFTA (CONICET, UNLP), Divisi\'on Qu\'imica Te\'orica\\
Blvd. 113 S/N, Sucursal 4, Casilla de Correo 16, 1900 La Plata, Argentina}
\maketitle

\begin{abstract}
We provide a pedagogical approach to the problem of avoided crossings
between electronic molecular curves and to diabatic and adiabatic
transitions when the nuclei of a diatomic molecule move according to
classical mechanics. For simplicity we restrict the analysis to only two
electronic states.
\end{abstract}

\section{Introduction}

\label{sec:intro}

Several physical and chemical phenomena are commonly described in terms of
the transition between adiabatic electronic curves that exhibit avoided
crossings. Such models are useful, for example, in the study of the
quenching of molecular species in electronically excited states by molecular
gases\cite{BFG69}, in photoinduced chemical reactions\cite{VO69,D75}, as
well as in the Marcus theory for electrochemical reactions\cite{BF01}. The
transition between the adiabatic curves is typically described by means of
the Landau-Zener formula\cite{Z32,RKLK81,W05}. This celebrated formula has
been tested by means of simple models for pedagogical purposes\cite{GA05}.
In addition to these applications to physical chemistry and molecular
physics we can also mention several models in classical mechanics developed
as pedagogical illustrative examples\cite{MQ88,FVB94,SGYR09,N10}. These
classical models are useful to test the main assumptions of the approach by
means of suitable devices that may be constructed in the laboratory\cite
{MQ88,FVB94,SGYR09}.

The purpose of this paper is to provide an additional pedagogical analysis
of the problem of avoided crossings and the adiabatic and diabatic
transitions between electronic states. In section~\ref{sec:BO} we outline
the Born-Oppenheimer approximation that is the source of the appearance of
the potential-energy surfaces in the quantum-mechanical treatment of
molecules. In section~\ref{sec:AC} we describe the avoided crossing between
two potential-energy curves and show some illustrative results provided by a
simple toy model. In section~\ref{sec:example} we analyse earlier
discussions on the avoided crossing between polar and nonpolar curves in
alkali halides. In section~\ref{sec:TDSchro} we briefly describe a
semiclassical approach in which the nuclei move according to classical
mechanics while the electronic states are treated by means of the
time-dependent Schr\"{o}dinger equation and illustrate the adiabatic and
diabatic transitions between an initial and a final state. Finally, in
section~\ref{sec:conclusions} we summarize the main results and draw
conclusions.

\section{The Born-Oppenheimer approximation}

\label{sec:BO}

The non-relativistic, time-independent Schr\"{o}dinger equation for a
diatomic molecule with $N$ electrons is
\begin{eqnarray}
H\psi &=&E\psi ,  \nonumber \\
H &=&T_{n}+T_{e}+V_{ne}+V_{ee}+V_{nn},  \label{eq:Schro}
\end{eqnarray}
where the kinetic energy of the nuclei $T_{n}$, the kinetic energy of the
electrons $T_{e}$ and the electron-nuclei, electron-electron and
nucleus-nucleus interactions $V_{ne}$, $V_{ee}$ and $V_{nn}$, respectively,
are given by
\begin{eqnarray}
T_{n} &=&-\frac{\hbar ^{2}}{2}\left( \frac{\nabla _{A}^{2}}{M_{A}}+\frac{%
\nabla _{B}^{2}}{M_{B}}\right) ,  \nonumber \\
T_{e} &=&-\frac{\hbar ^{2}}{2m}\sum_{j=1}^{N}\nabla _{j}^{2},  \nonumber \\
V_{ne} &=&-\frac{e^{2}}{4\pi \epsilon _{0}}\sum_{j=1}^{N}\left( \frac{Z_{A}}{%
r_{Aj}}+\frac{Z_{B}}{r_{Bj}}\right) ,  \nonumber \\
V_{ee} &=&\frac{e^{2}}{4\pi \epsilon _{0}}\sum_{i=1}^{N-1}\sum_{j=i+1}^{N}%
\frac{1}{r_{ij}},  \nonumber \\
V_{nn} &=&\frac{Z_{A}Z_{B}e^{2}}{4\pi \epsilon _{0}r_{AB}}.
\label{eq:T_V_etc}
\end{eqnarray}
In this expression $M_{A}$ and $M_{B}$ are the masses of the nuclei $A$ and $%
B$, respectively, $m$ is the electron mass, $Z_{A}$ and $Z_{B}$ are the
atomic numbers, $r_{Aj}$ and $r_{Bj}$ are the distances of the electron $j$
to each nucleus, $r_{ij}$ is the distance between electrons $i$ and $j$, $%
r_{AB}=R$ is the distance between the nuclei and $\nabla ^{2}$ denotes the
Laplacian for every kind of particle.

The Born-Oppenheimer approximation consists in writing the eigenfunctions
approximately as\cite{BH54,FE10}
\begin{equation}
\psi (\mathbf{r}_{j},R)\approx \psi ^{e}(\mathbf{r}_{j};R)\psi ^{n}(R),
\label{eq:psi_BO}
\end{equation}
where $\mathbf{r}_{j}$ stands for all the electron coordinates and the
electronic and nuclear functions $\psi ^{e}(\mathbf{r}_{j};R)$ and $\psi
^{n}(R)$, respectively, are solutions to
\begin{eqnarray}
H_{e}\psi ^{e} &=&E^{e}(R)\psi ^{e},  \nonumber \\
H_{e} &=&T_{e}+V_{ne}+V_{ee},  \label{eq:Schro_elect}
\end{eqnarray}
and
\begin{eqnarray}
H_{n}\psi ^{n} &=&E^{BO}\psi ^{n},  \nonumber \\
H_{n} &=&T_{n}+U(R),  \nonumber \\
U(R) &=&E^{e}(R)+V_{nn}(R).  \label{eq:Schro_nucl}
\end{eqnarray}
In this expression the Born-Oppenheimer energy $E^{BO}$ is the approximation
to the actual molecular energy $E$.

In this analysis we have omitted the separation of the motion of the center
of mass from the internal degrees of freedom that can be carried out in
equation (\ref{eq:Schro_nucl}) or, more rigorously, in equation (\ref
{eq:Schro})\cite{FE10}. In what follows we are interested in the
clamped-nuclei equation (\ref{eq:Schro_elect}) and such separation is not so
relevant.

\section{Avoided crossings}

\label{sec:AC}

In the rest of the paper we restrict ourselves to the electronic equation (%
\ref{eq:Schro_elect}) and for that reason we can omit the label $e$ on the
electronic Hamiltonian, its eigenfunctions and eigenvalues, without causing
(hopefully) any confusion. We suppose that we can obtain suitable
approximations to a pair of electronic states by means of the Rayleigh-Ritz
variational method with the simple trial function
\begin{equation}
\psi =c_{1}\varphi _{1}+c_{2}\varphi _{2},  \label{eq:psi_RR}
\end{equation}
where $\varphi _{1}$ and $\varphi _{2}$ are two suitable orthonormal
functions and $c_{1}$ and $c_{2}$ are variational coefficients. These
coefficients and the variational energies are solutions to the secular
equations
\begin{eqnarray}
\left( H_{11}-E\right) c_{1}+H_{12}c_{2} &=&0,  \nonumber \\
H_{21}c_{1}+\left( H_{22}-E\right) c_{2} &=&0,  \nonumber \\
H_{ij}=\left\langle \varphi _{i}\right| H\left| \varphi _{j}\right\rangle
&=&H_{ji}^{*}.  \label{eq:secular_eqs}
\end{eqnarray}
We obtain nontrivial solutions if $E$ is a root of the secular determinant.
The two roots of the characteristic polynomial
\begin{eqnarray}
E_{1} &=&\frac{H_{11}+H_{22}}{2}-\frac{\Delta }{2},  \nonumber \\
E_{2} &=&\frac{H_{11}+H_{22}}{2}+\frac{\Delta }{2},  \nonumber \\
\Delta &=&\sqrt{\left( H_{11}-H_{22}\right) ^{2}+4\left| H_{12}\right| ^{2}}.
\label{eq:E_RR}
\end{eqnarray}
lead to the two approximate electronic states
\begin{eqnarray}
\psi _{1} &=&c_{11}\varphi _{1}+c_{21}\varphi _{2},\;\frac{c_{21}}{c_{11}}=%
\frac{E_{1}-H_{11}}{H_{12}},  \nonumber \\
\psi _{2} &=&c_{12}\varphi _{1}+c_{22}\varphi _{2},\;\frac{c_{22}}{c_{12}}=%
\frac{E_{2}-H_{11}}{H_{12}},  \label{eq:psi_RR_j}
\end{eqnarray}
where $\left| c_{1j}\right| ^{2}+\left| c_{2j}\right| ^{2}=1$.

The matrix elements $H_{ij}$ depend on the internuclear distance $R$ and we
are interested in the case that $H_{11}(R)$ and $H_{22}(R)$ cross at $%
R=R_{c} $. For concreteness we assume that $H_{11}(R)<H_{22}(R)$ when $%
R<R_{c}$ and $H_{11}(R)>H_{22}(R)$ when $R>R_{c}$. If $H_{12}(R_{c})=0$ the
adiabatic energies $E_{1}(R)$ and $E_{2}(R)$ cross at $R_{c}$ but if $%
H_{12}(R_{c})\neq 0$ we are in the presence of an avoided crossing. In the
latter case the adiabatic energies approach each other as $R$ approaches $%
R_{c}$ and then move away as if repelling each other. At every $R$ the
energy difference is $E_{2}(R)-E_{1}(R)=\Delta (R)$ and reaches its minimum
at $R=R_{c}$, where $\Delta =2\left| H_{12}(R_{c})\right| $. At this point
\begin{eqnarray}
\psi _{1} &=&\frac{1}{\sqrt{2}}\left( \varphi _{1}-\frac{\left|
H_{12}\right| }{H_{12}}\varphi _{2}\right) ,  \nonumber \\
\psi _{2} &=&\frac{1}{\sqrt{2}}\left( \varphi _{1}+\frac{\left|
H_{12}\right| }{H_{12}}\varphi _{2}\right) .
\end{eqnarray}

Another common assumption is that $\left| E_{2}(R)-E_{1}(R)\right| \gg
\left| H_{12}(R)\right| $ if $\left| R-R_{c}\right| $ is sufficiently large;
therefore $\Delta \approx \left| H_{11}-H_{22}\right| $ under such
condition. For $R\ll R_{c}$ $E_{1}\approx H_{11}$ and $E_{2}\approx H_{22}$;
consequently $\psi _{1}\approx \varphi _{1}$ and $\psi _{2}\approx \varphi
_{2}$. On the other hand, if $R\gg R_{c}$ $E_{1}\approx H_{22}$, $%
E_{2}\approx H_{11}$, $\psi _{1}\approx \varphi _{2}$ and $\psi _{2}\approx
\varphi _{1}$. If $\varphi _{1}$ and $\varphi _{2}$ are associated to
different physical behaviours of the system (for example, a polar or a
nonpolar bond) we conclude that $\psi _{1}$ and $\psi _{2}$ change
considerably when the system goes from $R\ll R_{c}$ to $R\gg R_{c}$ along an
adiabatic curve.

\begin{figure}[tbp]
\begin{center}
\includegraphics[width=6cm]{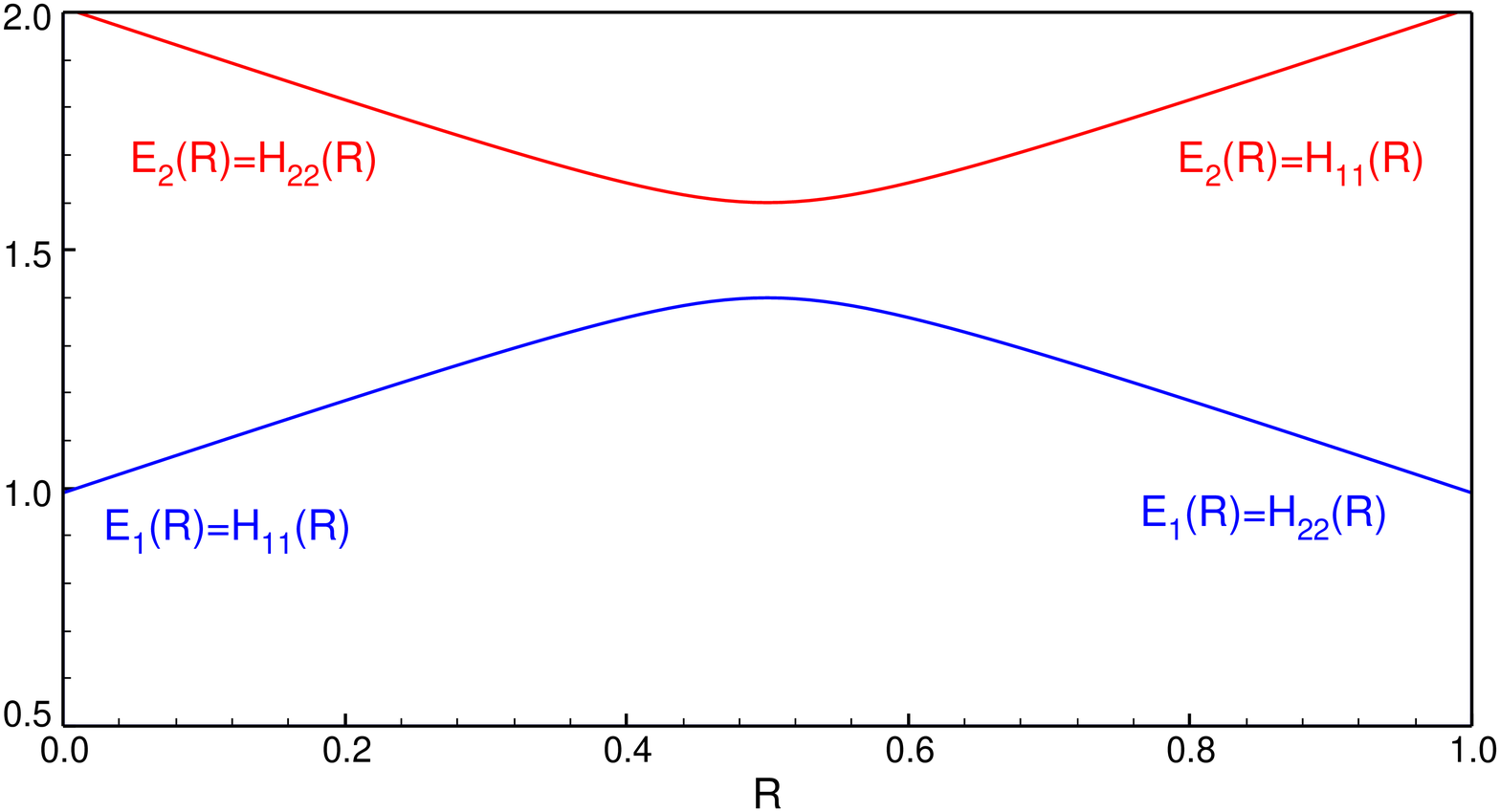}
\includegraphics[width=6cm]{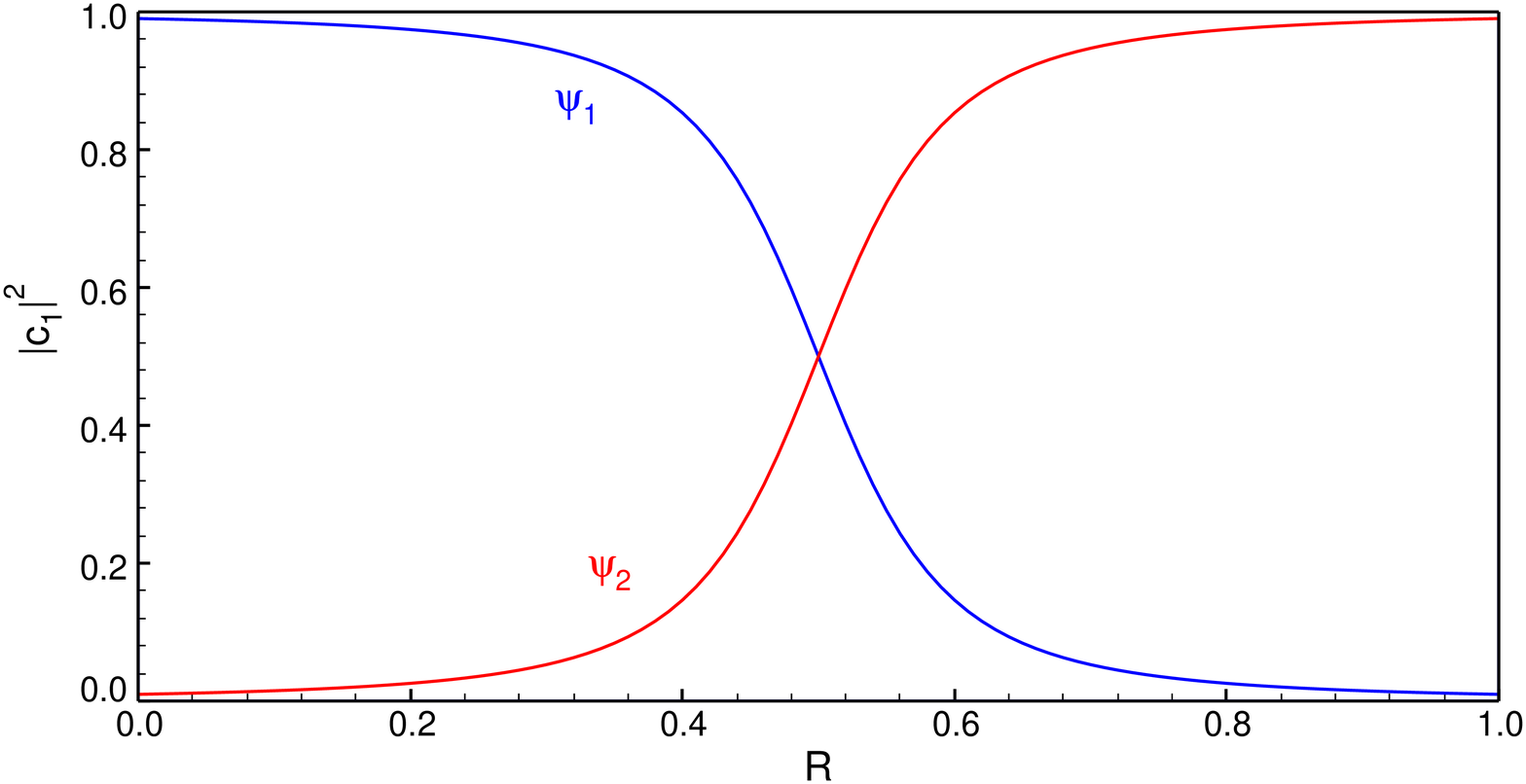}
\end{center}
\caption{Approximate energies (left panel) of the two-level model shown in
equations (\ref{eq:psi_RR}-\ref{eq:psi_RR_j}) and the contribution $%
\left|c_1\right|^2$ of $\varphi_1$ to the electronic states (right panel)}
\label{Fig:AC}
\end{figure}

Figure~\ref{Fig:AC} shows results for a toy model given by $H_{11}(R)=1+R$, $%
H_{22}(R)=2-R$ and $H_{12}(R)=0.1$. The energies in the left panel and the
square of the coefficient $\left| c_{1}(R)\right| ^{2}$ in the right one
clearly illustrate what we have just said. If we move from left to right
along the lower curve in figure~\ref{Fig:AC} left, we start in the state $%
\psi _{1}\approx \varphi _{1}$ and end in the state $\psi _{1}\approx
\varphi _{2}$. On the other hand, if we follow the upper curve we start in
the state $\psi _{2}\approx \varphi _{2}$ and end in the state $\psi
_{2}\approx \varphi _{1}$. These are examples of adiabatic transitions in
which we remain in the same adiabatic state $\psi _{j}$. If, on the other
hand, we start in $\psi _{1}\approx \varphi _{1}$ to the left and go through
the crossing towards the upper curve and end in $\psi _{2}\approx \varphi
_{1}$ then we are in the presence of a diabatic transition. We will discuss
these two processes with somewhat more detail in section~\ref{sec:TDSchro}.

\section{Example}

\label{sec:example}

\begin{figure}[tbp]
\begin{center}
\bigskip\bigskip\bigskip \includegraphics[width=6cm]{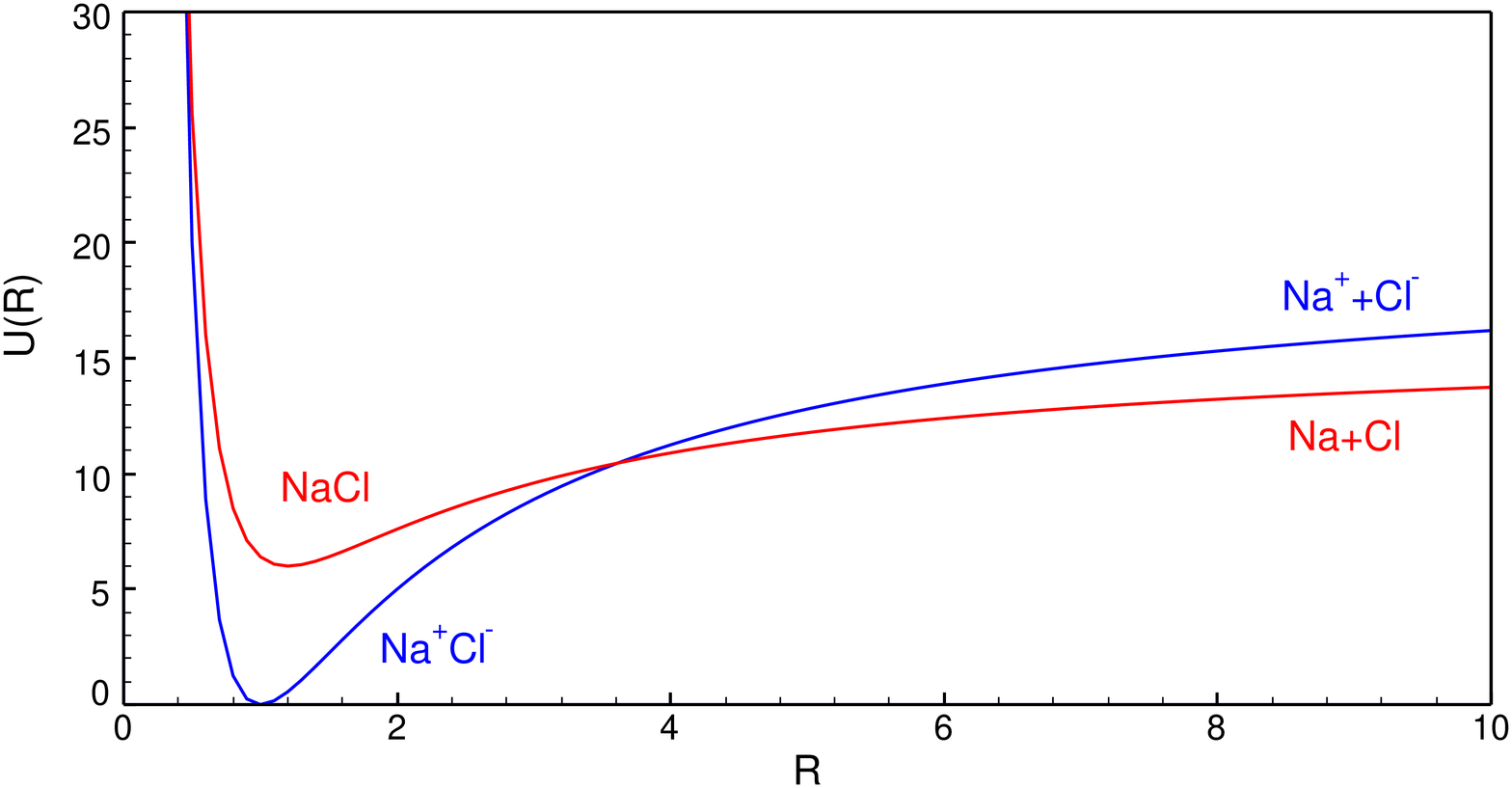}
\bigskip\bigskip\bigskip \includegraphics[width=6cm]{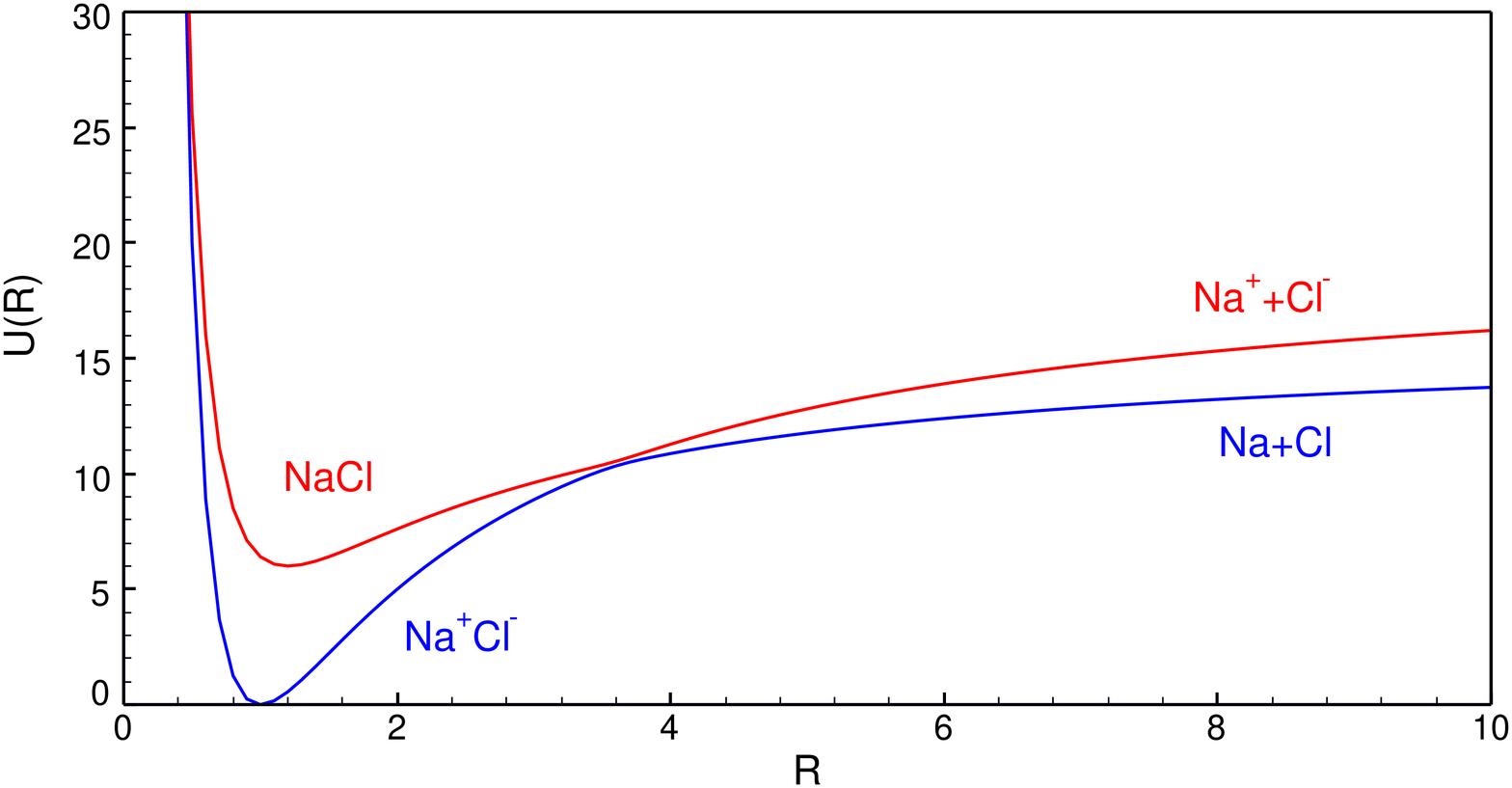}
\end{center}
\caption{Diabatic (left) and adiabatic (right) curves for NaCl }
\label{Fig:ACMX}
\end{figure}

Before proceeding with the general discussion of nonadiabatic transitions
let us consider the well known example of the alkali halides. For example,
Herzberg\cite{H50} mentions the avoided crossing between polar and nonpolar
potential energy curves of NaCl which are schematically shown in figure~\ref
{Fig:ACMX}. The left and right panels depict the diabatic and adiabatic
curves $U(R)=E_{e}(R)+V_{nn}(R)$, respectively. Note that in the latter case
the lowest electronic state changes from polar to nonpolar when going
through the avoided crossing from left to right while the upper state
behaves in the opposite way. The avoided crossing comes from the fact that
both states are $^{1}\Sigma ^{+}$ and cannot cross. Kauzmann\cite{K57} and
Devaquet\cite{D75} carried out an analysis in terms of diabatic functions of
the form
\begin{eqnarray}
\chi _{ion} &=&\phi _{Cl}(1)\phi _{Cl}(2)\left( \alpha _{1}\beta _{2}-\beta
_{1}\alpha _{2}\right) ,  \nonumber \\
\chi _{cov} &=&\left[ \phi _{Cl}(1)\phi _{Na}(2)+\phi _{Na}(1)\phi
_{Cl}(2)\right] \left( \alpha _{1}\beta _{2}-\beta _{1}\alpha _{2}\right) ,
\label{eq:chi-Cl-Na}
\end{eqnarray}
where $\phi _{Na}$ and $\phi _{Cl}$ are $3S$ and $3p$ atomic orbitals,
respectively.

According to Kauzmann: ``The true wave function of the NaCl molecule is a
hybrid of the above two
\begin{equation}
\psi =a\chi _{ion}+b\chi _{cov},  \label{eq:psi_ion_cov}
\end{equation}
the coefficients $a$ and $b$ being functions of the interatomic distance,
whose values, along with that of the corrected energy, may be found by means
of the Rayleigh-Ritz method.'' Devaquet\cite{D75} adds that ``the gap $g$
between the adiabatic states (in the case where the overlaps between $\chi
_{ion}$ and $\chi _{cov}$ is neglected) will be twice the matrix element $%
\left\langle \chi _{cov}\right| H\left| \chi _{ion}\right\rangle $ where $H$
denotes the total Hamiltonian of the molecule. Both one-electron and
two-electron terms in $H$ will give contributions.'' This analysis is
appealing but unfortunately one cannot take it seriously because the authors
failed to indicate why it is possible to describe the electronic structure
of a 28-electron molecule by means of a pair of two-electron wavefunctions.
One may reasonably ask about the meaning of the matrix element $\left\langle
\chi _{cov}\right| H\left| \chi _{ion}\right\rangle $ of a 28-electron
Hamiltonian between 2-electron functions. This kind of analysis should not
be carried out (even on a qualitative basis) unless one defines a suitable
model, which in this case means an effective two-electron Hamiltonian for an
approximate description of the two valence electrons. Alternatively, one
should indicate which 28-electron functions are schematically represented by
the two-electron functions (\ref{eq:psi_ion_cov}) that may probably be built
by means of the valence bond method.

\section{Time-evolution}

\label{sec:TDSchro}

In this section we discuss the time-evolution of an electronic molecular
state due to the classical motion of the two nuclei. According to quantum
mechanics the time-evolution of an state $\psi $ is given by the Schr\"{o}%
dinger equation
\begin{equation}
i\hbar \frac{\partial }{\partial t}\psi =H\psi .  \label{eq:Schro-TD}
\end{equation}
In order to derive general expressions it is convenient to consider an
ansatz of the form
\begin{equation}
\psi =\sum_{j}c_{j}(t)e^{-ie_{j}(t)/\hbar }\varphi _{j},  \label{eq:psi_TD}
\end{equation}
where $\left\{ \varphi _{j},\;j=1,2,\ldots \right\} $ is a complete set of
suitable orthonormal functions independent of time
\begin{equation}
\frac{\partial \varphi _{j}}{dt}=0.  \label{eq:dvarphi/dt=0}
\end{equation}
If $\psi $ is normalized at some initial time $t=t_{0}$ then it will be
normalized at all times because $H$ is Hermitian
\begin{equation}
\sum_{j}\left| c_{j}(t)\right| ^{2}=1.
\end{equation}
The quantities $e_{j}(t)$ depend on time, have units of
energy$\times $time and will be determined later. If we introduce
the ansatz (\ref{eq:psi_TD}) into (\ref{eq:Schro-TD}) we have
\begin{equation}
\sum_{j}\left( i\hbar \dot{c}_{j}+\dot{e}_{j}c_{j}\right) e^{-ie_{j}/\hbar
}\varphi _{j}=\sum_{j}c_{j}e^{-ie_{j}/\hbar }H\varphi _{j},
\end{equation}
where a point indicates time derivative. We now apply the ket $\left\langle
\varphi _{n}\right| $ from the left
\begin{equation}
\left( i\hbar \dot{c}_{n}+\dot{e}_{n}c_{n}\right) e^{-ie_{n}(t)/\hbar
}=\sum_{j}c_{j}e^{-ie_{j}/\hbar }H_{nj},\;H_{nj}=\left\langle \varphi
_{n}\right| H\left| \varphi _{j}\right\rangle ,\;n=1,2,\ldots ,
\end{equation}
and choose
\begin{equation}
\dot{e}_{n}=H_{nn},
\end{equation}
in order to remove the diagonal terms
\begin{equation}
i\hbar \dot{c}_{n}=\sum_{j\neq n}c_{j}e^{iu_{nj}/\hbar }H_{nj},\;u_{nj}=%
\frac{e_{n}-e_{j}}{\hbar }.  \label{eq:c_n-diff-eq}
\end{equation}
It is worth noting that the derivatives
\begin{equation}
\dot{u}_{nj}=\frac{H_{nn}-H_{jj}}{\hbar }=\omega _{nj},
\end{equation}
have units of angular frequency.

In order to apply these expressions to the two-level model discussed in
section~\ref{sec:AC} we restrict them to the case that $n,j=1,2$; therefore,
the system of equations (\ref{eq:c_n-diff-eq}) reduces to
\begin{eqnarray}
i\hbar \dot{c}_{1} &=&c_{2}e^{iu_{12}/\hbar }H_{12},  \nonumber \\
i\hbar \dot{c}_{2} &=&c_{1}e^{iu_{21}/\hbar }H_{21},
\end{eqnarray}
where $u_{12}=-u_{21}$ and $H_{12}=H_{21}^{*}$. For simplicity we define
\begin{equation}
w=\frac{H_{12}e^{iu_{12}}}{\hbar },
\end{equation}
that leads to somewhat simpler equations
\begin{eqnarray}
\dot{c}_{1} &=&-iwc_{2},  \nonumber \\
\dot{c}_{2} &=&-iw^{*}c_{1}.  \label{eq:c_i_coupled}
\end{eqnarray}

This problem is commonly analysed in terms of differential equations of
second order\cite{Z32,RKLK81,W05}. To this end we differentiate the first
equation in (\ref{eq:c_i_coupled}) with respect to time and then express $%
\dot{c}_{2}$ and $c_{2}$ in terms of $c_{1}$ and $\dot{c}_{1}$ using the
same system of equations:
\begin{equation}
\ddot{c}_{1}=-i\dot{w}c_{2}-iw\dot{c}_{2}=\frac{\dot{w}}{w}\dot{c}%
_{1}-\left| w\right| ^{2}c_{1}.  \label{eq:dif_eq_c_1}
\end{equation}
Analogously, we can derive a similar equation for $c_{2}$:
\begin{equation}
\ddot{c}_{2}=\frac{\dot{w}^{*}}{w^{*}}\dot{c}_{2}-\left| w\right| ^{2}c_{2}.
\end{equation}

If we assume that $\psi (t_{0})=\varphi _{1}$ (that is to say: $%
c_{1}(t_{0})=1$, $c_{2}(t_{0})=0$) then the initial conditions for the
differential equation (\ref{eq:dif_eq_c_1}) are
\begin{equation}
c_{1}(t_{0})=1,\;\dot{c}_{1}(t_{0})=0.  \label{eq:inic_cond_a}
\end{equation}
The probability that the system remains in $\varphi _{1}$ at some $t>t_{0}$
is given by
\begin{equation}
P_{1}(t)=\left| \left\langle \varphi _{1}\right| \left. \psi
(t)\right\rangle \right| ^{2}=\left| c_{1}(t)\right| ^{2}.
\end{equation}

As argued in the preceding section $H_{11}(R)$ and $H_{22}(R)$ cross at $%
R=R_{c}$. We can expand all the relevant quantities in a Taylor series about
this point:

\begin{eqnarray}
\varphi _{k}(r_{j},R) &=&\varphi _{k}(r_{j},R_{c})+\varphi
_{k}^{(1)}(r_{j},R_{c})\left( R-R_{c}\right) +\ldots .  \nonumber \\
H_{12}(R) &=&H_{12}(R_{c})+H_{12}^{(1)}(R_{c})\left( R-R_{c}\right) +\ldots .
\nonumber \\
H_{11}(R)-H_{22}(R) &=&\Delta H^{(1)}(R_{c})\left( R-R_{c}\right) +\ldots .
\label{eq:expansions_R-Rc}
\end{eqnarray}
If we just keep the leading terms we can assume that $\varphi _{k}$ and $%
H_{12}$ are almost independent of $R$ in such a first-order approximation
and that $H_{11}-H_{22}$ varies linearly with $R$. The coefficient
\begin{equation}
\Delta H^{(1)}(R_{c})=\left. \frac{\partial }{R}\left( H_{11}-H_{22}\right)
\right| _{R=R_{c}},  \label{eq:slope_Rc}
\end{equation}
will be relevant for subsequent discussion. Note that it is the difference
between the slopes of $H_{11}(R)$ and $H_{22}(R)$ at the diabatic crossing $%
R=R_{c}$.

Let us assume that the nuclei move according to classical mechanics with a
constant velocity $\dot{R}=v$ so that $R(t)=R_{0}+v\left( t-t_{0}\right) $.
If $R\left( t_{c}\right) =R_{c}$ we have $R_{c}=R_{0}+v\left(
t_{c}-t_{0}\right) $ and $R-R_{c}=v\left( t-t_{c}\right) $. Because of what
we have just argued about equations (\ref{eq:expansions_R-Rc}) we conclude
that the functions $\varphi _{k}$ are independent of time, which is
consistent with equation (\ref{eq:dvarphi/dt=0}), and that
\begin{equation}
\frac{dH_{12}}{dt}=0.  \label{eq:dH12/dt=0}
\end{equation}
For this reason
\begin{equation}
\dot{w}=i\dot{u}_{12}w=i\frac{H_{11}-H_{22}}{\hbar }w=i\omega _{12}w,
\end{equation}
and equation (\ref{eq:dif_eq_c_1}) becomes
\begin{equation}
\ddot{c}_{1}=i\omega _{12}\dot{c}_{1}-\left| w\right| ^{2}c_{1},\;\left|
w\right| =\frac{\left| H_{12}\right| }{\hbar }.  \label{eq:dif_eq_c_1_b}
\end{equation}
From all the assumptions outlined above we conclude that
\begin{equation}
\omega _{12}=\frac{H_{11}-H_{22}}{\hbar }=\frac{\alpha \left( t-t_{c}\right)
}{\hbar },\;\alpha =\Delta H^{(1)}v.  \label{eq:alpha_definition}
\end{equation}
Thus, the differential equation of second order becomes
\begin{equation}
\ddot{c}_{1}=i\frac{\alpha \left( t-t_{c}\right) }{\hbar }\dot{c}_{1}-\left|
w\right| ^{2}c_{1}.  \label{eq:dif_eq_c_1_c}
\end{equation}

All the calculations are much simpler if we work with dimensionless
equations with the smallest number of model parameters. To this end we
define the dimensionless time
\begin{equation}
s=\left| w\right| \left( t-t_{c}\right) =\left| \frac{H_{12}}{\hbar }\right|
\left( t-t_{c}\right) ,
\end{equation}
so that
\begin{equation}
c_{1}^{\prime \prime }=i\lambda sc_{1}^{\prime }-c_{1},\;\lambda =\frac{%
\alpha }{\hbar \left| w\right| ^{2}}=\frac{\hbar \alpha }{\left|
H_{12}\right| ^{2}}=\frac{\hbar \Delta H^{(1)}v}{\left| H_{12}\right| ^{2}},
\label{eq:dif_eq_c_1_d}
\end{equation}
where the prime stands for derivative with respect to $s$. Since $\dot{c}%
_{1}=\left| w\right| c_{1}^{\prime }$, the boundary conditions become
\begin{equation}
c_{1}(s_{0})=1,\;c_{1}^{\prime }(s_{0})=0.  \label{eq:inic_cond_b}
\end{equation}
The advantage of equation (\ref{eq:dif_eq_c_1_d}) is that it depends on only
one parameter $\lambda $. Note that it increases with the difference between
the slopes of $H_{11}(R)$ and $H_{22}(R)$ at $R=R_{c}$, with the relative
velocity of the nuclei and decreases with the magnitude of the gap between
the adiabatic curves at $R=R_{c}$. When $\lambda =0$ the system simply
oscillates between the two states: $c_{1}(s)=\cos \left( s-s_{0}\right) $.

\begin{figure}[tbp]
\begin{center}
\bigskip\bigskip\bigskip \includegraphics[width=9cm]{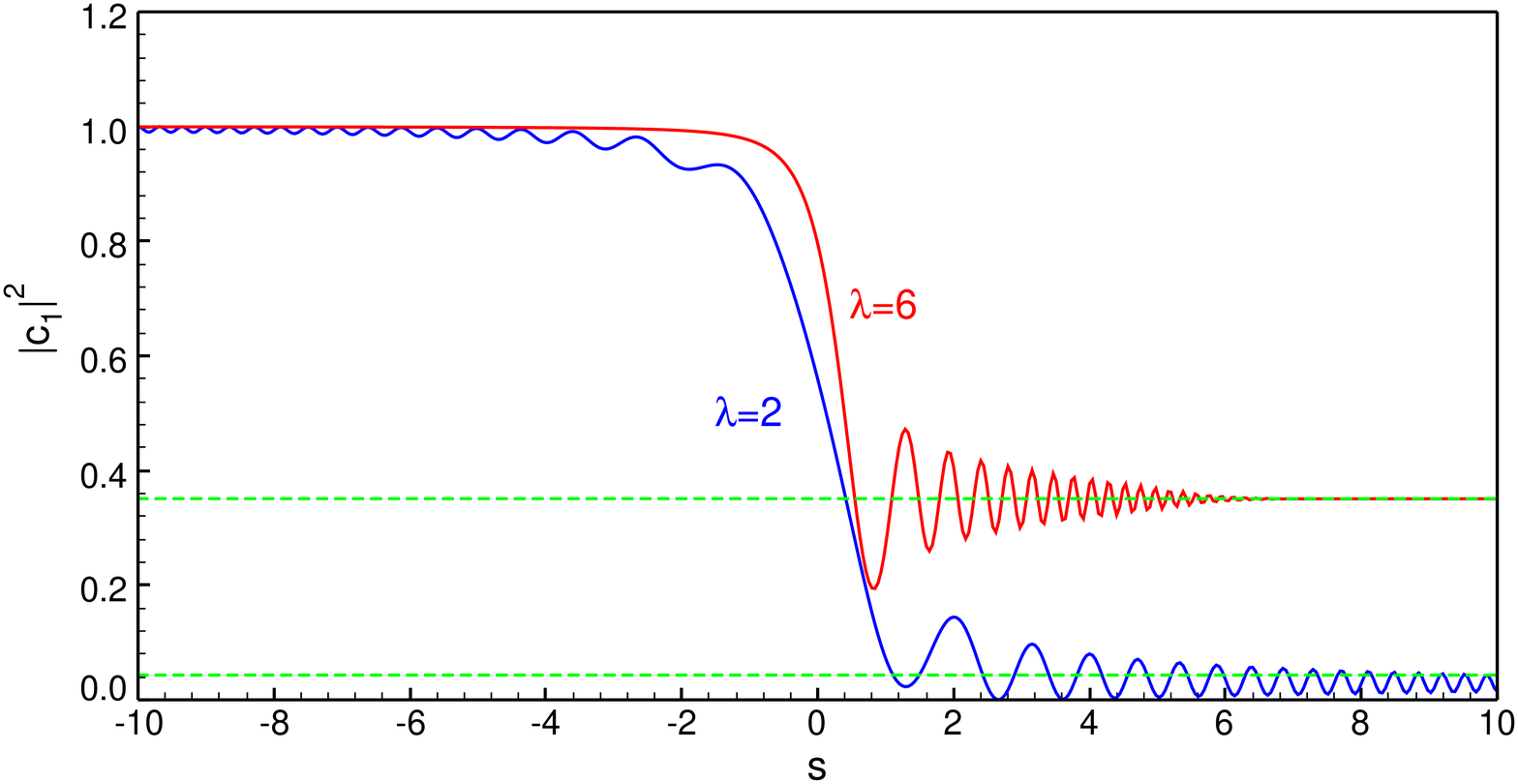}
\end{center}
\caption{Probability that $\psi(s)=\varphi_1$ for two values of $\lambda$.
The dashed (green) lines mark the Landau-Zener limits }
\label{Fig:LZprob}
\end{figure}

Figure~\ref{Fig:LZprob} shows $\left| c_{1}\right| ^{2}(s)$ for $s_{0}=-10$
and two values of $\lambda $. We appreciate that the probability that the
system remains in the state $\varphi _{1}$ when $s\rightarrow \infty $
agrees with the limit given by the celebrated Landau-Zener formula\cite
{Z32,RKLK81,W05}
\begin{equation}
P_{1}=\exp \left( -\frac{2\pi }{\lambda }\right) ,
\end{equation}
written in terms of the only parameter in the dimensionless equation (\ref
{eq:dif_eq_c_1_d}). On the other hand, the probability of the transition
from the state $\varphi _{1}$ to the state $\varphi _{2}$ is given by $%
P_{2}=1-P_{1}$. Note that $\left| c_{2}\right| ^{2}(s)=1-$ $\left|
c_{1}\right| ^{2}(s)$ is the probability of a diabatic transition from the
lower to the upper curve in figure~\ref{Fig:AC} left. On the other hand, $%
\left| c_{1}\right| ^{2}(s)$ is the adiabatic transition in which the system
remains in the lower curve in that same figure. Figure~\ref{Fig:LZprob}
shows that the probability of an adiabatic transition increases with $%
\lambda $. We obtained the numerical data for this figure by means of the
fourth-order Runge-Kutta method built in Derive\cite{DERIVE96} (see page
266).

In closing this section we mention that the Landau-Zener approach exhibits
several limitations already summarized by Geltman and Aragon\cite{GA05} in a
pedagogical paper (more rigorous analyses can be seen in the references
therein).

\section{Conclusions}

\label{sec:conclusions}

We have seen that two adiabatic potential-energy curves cannot cross unless
their interaction vanishes at the crossing point. Commonly, they exhibit an
avoided crossing that looks as if they repel each other. If the nuclei are
treated as classical particles they can remain on the same adiabatic curve
or move from one to the other (adiabatic and diabatic transitions,
respectively). The probability of each of these processes is determined by
the relative velocity of the nuclei, the slopes of the diabatic energies at
the crossing and the gap between the adiabatic curves at such point. The
process is described by a differential equation of second order that leads
to the celebrated Landau-Zener formula for the transition probability\cite
{Z32,RKLK81,W05}. There is a trick in the development of such differential
equation: the functions $\varphi _{j}$ and the interaction $H_{12}$ are
assumed to be time-independent while $H_{11}-H_{22}$ is time-dependent. That
this approximation is sound can be verified experimentally by means of
classical devices\cite{MQ88,FVB94,SGYR09}.

It is worth noting that the potential-energy curves (or, more generally, the
potential-energy surfaces) appear in a quantum-mechanical description of
molecular systems because of the application of the Born-Oppenheimer
approximation\cite{BH54,FE10}. In this sense, we may say that the
potential-energy surfaces are artifacts of such an approximation. It may be
interesting to investigate how to obtain results and conclusions similar to
those in the preceding sections without that approximation. In other words,
how to describe the phenomena outlined above by means of the time-dependent
Schr\"{o}dinger equation with the complete Hamiltonian shown in equation (%
\ref{eq:Schro}).

\section*{Acknowledgments}

The author would like to thank Adela Croce and Waldemar Marmisoll\'e for
bringing this problem to his attention.

\begin{thebibliography}{99}
\bibitem{BFG69}  E. Bauer, E. R. Fisher, and F. R. Gilmore, De-excitation of
Electronically Excited Sodium by Nitrogen, J. Chem. Phys. 51 (1969)
4173-4181.

\bibitem{VO69}  W. Th. A. M. Van der Lugt and L. J. Oosterhoff, Symmetry
control and photoinduced reactions, J. Amer. Chem. Soc. 91 (1969) 6042-6049.

\bibitem{D75}  A. Devaquet, Avoided crossings in photochemistry, Pure Appl.
Chem. 41 (1975) 455-473.

\bibitem{BF01}  A. J. Bard and L. R. Faulkner, Electrochemical Methods.
Fundamentals and Applications, Second (John Wiley \& Sons, New York, 2001).

\bibitem{Z32}  C. Zener, Non-Adiabatic Crossing of Energy Levels, Proc. R.
Soc. London Ser. A137 (1932) 696-702.

\bibitem{RKLK81}  J. R. Rubbmark, M. M. Kash, M. G. Littman, and D.
Kleppner, Dynamical effects at avoided level crossings: A study of the
Landay-Zerner effect using Rydberg atoms, Phys. Rev. A 23 (1981) 3107-3117.

\bibitem{W05}  C. Wittig, The Landau-Zener Formula, J. Phys. Chem. B 109
(2005) 8428-8430.

\bibitem{GA05}  S. Geltman and N. D. Aaragon, Model study of the
Landau-Zener approximation, Am. J. Phys. 73 (2005) 1050-1054.

\bibitem{MQ88}  H. J. Maris and X. Quan, Adiabatic and nonadiabatic
processes in classical and quantum mechanics, Am. J. Phys. 56 (1988)
1114-1117.

\bibitem{FVB94}  W. Frank and P. von Brentano, Classical analogy to quantum
mechanical level repulsion, Am. J. Phys. 62 (1994) 706-709.

\bibitem{SGYR09}  B. W. Shore, M. V. Gromovyy, L. P. Yatsenko, and V. I.
Romanenko, Simple mechanical analogs of rapid adiabatic passage in atomic
physics, Am. J. Phys. 77 (2009) 1183-1194.

\bibitem{N10}  L. Novotny, Strong coupling, energy splitting, and level
crossings: A classical perspective, Am. J. Phys. 78 (2010) 1199-1202.

\bibitem{BH54}  M. Born and K. Huang, Dynamical Theory of Cristal Lattices,
Oxford University Press, Glasgow, 1954).

\bibitem{FE10}  F. M. Fern\'{a}ndez and J. Echave, Nonadiabatic Calculation
of Dipole Moments, in: J. Grunenberg (Ed.), Computational Spectroscopy.
Methods, Experiments and Applications, Vol. Wiley-VCH, Weinheim, 2010.

\bibitem{H50}  G. Herzberg, Molecular Spectra and Molecular Structure. I.
Spectra of Diatomic Molecules, Second (Van Nostrand Reinhold Company, New
York, 1950).

\bibitem{K57}  W. Kauzmann, Quantum Chemistry, Academic Press, New York,
1957).

\bibitem{DERIVE96}  A. Rich, J. Rich, T. Shelby, and D. Stoutemyer, User
Manual Derive, Seventh (Soft Warehouse, Honolulu, 1996).
\end{thebibliography}
\end{document}